\documentclass[12pt]{article}

\usepackage{amssymb}
\usepackage{amsmath}

\usepackage{epsfig}

\begin{document} %

\title{\bf Randall-Sundrum model with a small curvature and dimuon production
at the LHC}

\author{A.V. Kisselev\thanks{Electronic address:
alexandre.kisselev@ihep.ru} \\
{\small Institute for High Energy Physics, 142281 Protvino, Russia}
\\
{\small and}
\\
{\small Department of Physics, Moscow State University, 119991
Moscow, Russia}}

\date{}

\maketitle

\thispagestyle{empty}


\begin{abstract}
The $p_{\perp}$-distributions for the dimuon  production at the LHC
are calculated in the Randall-Sundrum scenario with a small
curvature $\kappa$. The $pp$ collisions at 7 TeV and 14 TeV are
considered. The widths of massive graviton excitations are taken
into account. It is found that the LHC discovery limit on
5-dimensional gravity scale $M_5$ is equal to 5.5 TeV for $\sqrt{s}
= 7$ TeV and integrated luminositiy 5 fb$^{-1}$. For $\sqrt{s} = 14$
TeV and integrated luminosities 30 fb$^{-1}$ and 100 fb$^{-1}$, the
search limits in the dimuon events are 12.0 TeV and 14.6 TeV,
respectively. Contrary to the standard RS model, these limits are
independent of $\kappa$, provided $\kappa \ll M_5$.
\end{abstract}




\section{Extra dimension with the small curvature}
\label{sec:1}

The goal of the present paper is to estimate gravity effects in the
dimuon production at the LHC in a scheme with a warp extra
dimension, and to derive search limits on a fundamental
5-dimensional gravity scale $M_5$.

The standard Randall-Sundrum (RS) model~\cite{Randall:99} is the
theory with one extra dimension (ED) in a slice of the AdS$_5$
space-time. It predicts a series of massive Kaluza-Klein (KK)
resonances with the lightest one, $m_1$, around one TeV. A lot of
efforts were made in order to find effects coming from warped ED.
The bounds on 5-dimensional Planck scale $M_5$ and/or $m_1$ were
obtained both at the Tevatron~\cite{RS:Tevatron} and recently at the
LHC. In particular, high-mass dilepton~\cite{RS:dilepton_res},
diphoton~\cite{RS:diphoton_res} and $t\bar{t}$ resonance
events~\cite{RS:ttbar_res} were searched for in $pp$ collisions at
$\sqrt{s} = 7$ TeV with the ATLAS and CMS experiments (see also
\cite{RS:dileptons_event_sel}, \cite{RS:high_mass_res}).

In the present paper we will study the RS scenario with the
5-dimensional Planck scale $M_5$ in the TeV region and \emph{small
curvature}~\cite{Kisselev:05}-\cite{Kisselev:diphotons}:%
\footnote{The ratio of the 5-dimensional scalar curvatures in our
scheme $\mathcal{R}$ to the scalar curvature in the standard RS
scheme $\mathcal{R}_0$ is given by $\mathcal{R}/\mathcal{R}_0 \simeq
(\kappa/\bar{M}_5)^3 \ll 1$.}
\begin{equation}\label{scale_relation}
\kappa \ll M_5 \;.
\end{equation}
In such a scheme, the background warped metric looks like
\begin{equation}\label{metric}
ds^2 = e^{2 \kappa (\pi r_c - |y|)} \, \eta_{\mu \nu} \, dx^{\mu} \,
dx^{\nu} + dy^2 \;.
\end{equation}
Here $y = r_c \, \theta$ ($-\pi \leq \theta \leq \pi$) is the 5-th
dimension coordinate, $r_c$ is the size of the ED, $\eta_{\mu \nu}$
is the Minkowski metric. The points $(x_{\mu}, y)$ and $(x_{\mu},
-y)$ are identified, so one gets the orbifold
$\mathrm{S}^1/\mathrm{Z_2}$. The parameter $\kappa$ defines the
5-dimensional scalar curvature of the AdS$_5$ space. In what
follows, we will call it \emph{curvature}.

We are interested in the RS scheme with two 3D branes with equal and
opposite tensions located in 5-th dimension at the points $y = \pi
r_c$ (called TeV brane) and  $y = 0$ (called Plank brane). The SM
fields are constrained to the TeV brane, while the gravity
propagates in all five spatial dimensions.

It is necessary to note that metric (\ref{metric}) is chosen in such
a way that 4-dimensional coordinates $x^{\mu}$ are Galilean on the
TeV brane where all the SM fields live, since the warp factor is
equal to unity at $y = \pi r_c$. Due to the warp factor in the
metric \eqref{metric}, the gravity is strong on the Planck brane,
while it is weak on the TeV brane.

By integrating 5-dimensional action in variable $y$, one can get an
effective 4-dimensional action which, in its turn, leads to
so-called \emph{hierarchy relation} between the 5-dimensional
\emph{reduced} gravity scale $\bar{M}_5$ and \emph{reduced} Planck
mass $\bar{M}_{\mathrm{Pl}}$:
\begin{equation}\label{RS_hierarchy_relation}
\bar{M}_{\mathrm{Pl}}^2 = \frac{\bar{M}_5^3}{\kappa} \left( e^{2 \pi
\kappa r_c} - 1 \right) \;.
\end{equation}
The fundamental gravity scale $M_5$ (i.e., Planck scale in five
dimensions) and reduced scale $\bar{M}_5$ are related as follows:
\begin{equation}\label{reduced_scale}
M_5 = (2\pi)^{1/3} \bar{M}_5 \simeq 1.84 \, \bar{M}_5 \;.
\end{equation}
In order the hierarchy relation (\ref{RS_hierarchy_relation}) to be
satisfied, it is enough to take $r_c \kappa \simeq 8 \div 9.5$, that
corresponds to $r_c \simeq 0.15 \div 1.8 \mathrm{\ fm}$. In what
follows, we choose $\bar{M}_5$ to be of the order of several TeV to
tens TeV, while $\kappa$ is allowed to vary from hundred MeV to tens
GeV. Thus, no new parameters of the order of $\bar{M}_{\mathrm{Pl}}$
(as in the standard RS model \cite{Randall:99}, in which $\kappa
\sim M_5 \sim M_{\mathrm{Pl}}$) are introduced in our scheme.

From the point of view of a 4-dimensional observer located on the
TeV brane, in addition to the massless graviton, there exists an
infinite number of its Kaluza-Klein (KK) excitations, $G^{(n)}_{\mu
\nu}$, with the masses
\begin{equation}\label{graviton_masses}
m_n = x_n \, \kappa, \qquad n=1,2 \ldots \;,
\end{equation}
where $x_n$ are zeros of the Bessel function $J_1(x)$, and $n$ is
the KK-number.

The interaction of the KK gravitons with the the SM fields on the
TeV brane is described by the Lagrangian
\begin{equation}\label{Lagrangian}
\mathcal{L}_{int} = - \frac{1}{\bar{M}_{\mathrm{Pl}}} \, T^{\mu \nu}
\, G^{(0)}_{\mu \nu} - \frac{1}{\Lambda_{\pi}} \, T^{\mu \nu} \,
\sum_{n=1}^{\infty} G^{(n)}_{\mu \nu} \;,
\end{equation}
were $T^{\mu \nu}$ is the energy-momentum tensor of the SM fields.
The parameter
\begin{equation}\label{lambda}
\Lambda_{\pi} = \bar{M}_5 \,\left( \frac{\bar{M}_5}{\kappa}
\right)^{\! 1/2}
\end{equation}
can be regarded as a physical scale on the TeV brane where all SM
fields live.

The scalar field, radion, is omitted in the Lagrangian
\eqref{Lagrangian}. In our scheme, it practically decouples from the
SM fields since its coupling on the TeV brane is proportional
to~\cite{Kisselev:05}
\begin{equation}\label{radion_coupling}
g_{\mathrm{rad}}^2 \simeq 3 \cdot 10^{-4} \left(
\frac{\kappa}{\mathrm{GeV}} \right) \left(
\frac{\mathrm{TeV}}{\bar{M}_5} \right)^3 \mathrm{TeV}^{-2} \;.
\end{equation}
In particular, for $\kappa = 1$ GeV and $\bar{M}_5 = 1$ TeV, one
gets $g_{\mathrm{rad}} \sim 1/(100 \ \mathrm{TeV})$.

As for processes with a real or virtual production of the KK
gravitons, the smallness of their coupling to the SM particles,
$1/\Lambda_{\pi}$, is compensated by a large number of real
gravitons or by infinite tower of virtual KK excitations. As a
result, a magnitude of the cross sections is defined by the scale
$\bar{M}_5$, not the scale $\Lambda_{\pi}$~\cite{Kisselev:05} (as an
illustration, see \eqref{gravity_cs_1}).

As it was already mentioned above, the 4-dimensional coordinates are
Galilean on the TeV brane. Thus, a correct determination of the
masses on this brane can be achieved~\cite{Boos:02}. In particular,
observed masses coincide with their Lagrangian values and do not
depend on a coordinate rescaling, if covariant equations and
invariant distances are used~\cite{Grinstein:01}.

The RS-like model under consideration has other interesting
features. For instance, the spectrum of the KK gravitons
\eqref{graviton_masses} is very similar to that in the ADD model
with one ED~\cite{Arkani-Hamed:98}. Moreover, all matrix elements
can be formally obtained from corresponding matrix elements
calculated in the ADD model with one \emph{flat} extra dimension by
using following replacements~\cite{Kisselev:05, Kisselev:06}:
\begin{equation}\label{RS_vs_ADD}
\bar{M}_{4+1} \rightarrow (2 \pi)^{-1/3} \, \bar{M}_5 \;, \qquad R_c
\rightarrow (\pi \kappa)^{-1} \;.
\end{equation}
Here $\bar{M}_{4+1}$ is a 5-dimensional reduced Planck scale, $R_c$
being the radius of the compact flat dimension.

\section{Gravitons in dimuon production at the LHC}
\label{sec:2}

In this section we will study a production of two muons with high
transverse momenta in $pp$-collisions~\cite{Matveev:69}:%
\footnote{The production of a lepton pair in $pp$ (or $p\bar{p}$)
collision is often called Drell-Yan process.}
\begin{equation}\label{process}
p \, p \rightarrow \mu^+ \mu^- + X \;.
\end{equation}
We are interested in a possible excess in $p_{\perp}$-distribution
of the final leptons coming from gravity interactions in
5-dimensional space-time with the small curvature. The theoretical
framework was described in details in the previous section.

The dilepton production is a favourable process to look for gravity
effects predicted by theories based on a space-time with more than
four dimensions. The Tevatron data on $p_{\perp}$-distribution in
dilepton production already provided us with bounds on parameters of
these theories~\cite{DY:Tevatron}. As for dilepton production at the
LHC, the search for EDs resulted in new limits on so-called large
EDs \cite{LHC_limits:ADD}. However, most efforts of ATLAS and CMS
Collaborations were made to search for massive resonances in such
events predicted by theories with the warped metric (see, for
instance, \cite{RS:dilepton_res,RS:dileptons_event_sel}). The
phenomenology of the RS model with the resonant KK spectrum in the
TeV region can be found in \cite{Davoudiasl:00}. The determination
of the spin-2 RS gravitons against spin-1 and spin-0 resonances in
lepton-pair production was done in \cite{Osland:08}.

The goal of this paper is to estimate gravity effects in the dimuon
production at the LHC  in the RS scenario with the small
curvature~\cite{Kisselev:05}-\cite{Kisselev:06}. The differential
cross section of this process \eqref{process} is given by
\begin{align}\label{cross_sec}
\frac{d \sigma}{d p_{\perp}}(p p \rightarrow  \mu^+ \mu^- + X) &=
2p_{\perp} \!\!\!\! \sum\limits_{a,b = q,\bar{q},g} \!\!
\int\nolimits \!\! \frac{d\tau \sqrt{\tau}}{\sqrt{\tau -
x_{\perp}^2}} \! \int\nolimits \! \frac{dx_1}{x_1}  f_{a/p}(\mu^2,
x_1)
\nonumber \\
&\times f_{b/p}(\mu^2, \tau/x_1) \, \frac{d
\hat{\sigma}}{d\hat{t}}(a b \rightarrow  \mu^+ \mu^-) \;,
\end{align}
with the transverse energy of the muon pair equals to $2p_{\perp}$.
Here $ f_{a/p}(\mu^2, x_1)$ is the distribution of the parton of the
type $a$ in momentum fraction $x_1$ inside the proton taken at the
scale $\mu$. $d\hat{\sigma}/d\hat{t}$ denotes the cross section of
the partonic subprocess $a b \rightarrow \mu^+ \mu^-$, which is
described by the Mandelstam variables $\hat{s}$, $\hat{t}$ and
$\hat{u}$ ($\hat{s} + \hat{t} + \hat{u} = 0$).

We introduced two dimensionless quantities,
\begin{align}\label{tau_xtr}
x_{\perp} &= \frac{2 p_{\perp}}{\sqrt{s}} \;,
\nonumber \\
\tau &= x_1 x_2 \,,
\end{align}
where $x_2$ is the momentum fraction of the parton $b$ in
\eqref{cross_sec}. Without cuts, integration variables in
\eqref{cross_sec} vary within the following limits
\begin{align}\label{int_region_full}
x_{\perp}^2 & \leq \tau \leq 1 \;, \nonumber \\
\tau & \leq x_1 \leq 1 \;.
\end{align}
After imposing kinematical cut, the integration region becomes more
complicated (see Appendix~A).

The contribution of the virtual gravitons to lepton pair production
($l = e \mathrm{\ or \ } \mu$) comes from the quark-antiquark
annihilation, $q \, \bar{q} \rightarrow G^{(n)} \rightarrow l^+l^-$,
and gluon-gluon fusion, $g \, g \rightarrow G^{(n)} \rightarrow
l^+l^-$. The corresponding partonic cross sections are (see, for
instance, Appendix~A in \cite{Giudice:05})
\begin{eqnarray} \label{parton_cross_sec}
\frac{d\hat{\sigma}}{d\hat{t}}(q \bar{q} \rightarrow l^+l^-) &=&
\frac{\hat{s}^4 + 10\hat{s}^3 \hat{t} + 42 \, \hat{s}^2 \hat{t}^2 +
64 \hat{s} \, \hat{t}^3 + 32 \, \hat{t}^4}{1536 \, \pi \hat{s}^2}
\left| \mathcal{S}(\hat{s}) \right|^2 \;,
\nonumber \\
\frac{d\hat{\sigma}}{d\hat{t}}(gg \rightarrow l^+l^-) &=&
-\frac{\hat{t}(\hat{s} + \hat{t}) (\hat{s}^2 + 2 \hat{s}\,\hat{t} +
2\,\hat{t}^2)}{256 \, \pi \hat{s}^2} \left|\mathcal{S}(\hat{s})
\right|^2
 \;,
\end{eqnarray}
where
\begin{equation}\label{KK_summation}
\mathcal{S}(s) = \frac{1}{\Lambda_{\pi}^2} \sum_{n=1}^{\infty}
\frac{1}{s - m_n^2 + i \, m_n \Gamma_n} \;
\end{equation}
is the invariant part of the partonic matrix elements, with
$\Gamma_n$ being total width of the graviton with the KK number $n$
and mass $m_n$~\cite{Kisselev:06}:
\begin{equation}\label{graviton_widths}
\Gamma_n = \eta \, m_n \left( \frac{m_n}{\Lambda_{\pi}} \right)^2,
\quad \eta \simeq 0.09 \;.
\end{equation}
Note, because of the universal coupling of the KK gravitons to all
SM fields, the function $\mathcal{S}(s)$ is the same for all
processes mediated by $s$-channel virtual gravitons.

In the RS scenario with the small curvature, sum
(\ref{KK_summation}) was calculated analytically in
\cite{Kisselev:06}. At $s \sim \bar{M}_5 \gg \kappa$, it looks like
\begin{equation}\label{KK_sum}
\mathcal{S}(s) = - \frac{1}{4 \bar{M}_5^3 \sqrt{s}} \; \frac{\sin 2A
+ i \sinh 2\varepsilon }{\cos^2 \! A + \sinh^2 \! \varepsilon} \;,
\end{equation}
where
\begin{equation}\label{parameters}
A = \frac{\sqrt{s}}{\kappa} \;, \qquad \varepsilon = \frac{\eta}{2}
\Big( \frac{\sqrt{s}}{\bar{M}_5} \Big)^3 \;.
\end{equation}

It is interesting to note that the magnitude of $\mathcal{S}(s)$ is
defined by the fundamental gravity scale $\bar{M}_5$, while the
scale $\Lambda_{\pi}$, which is contained in the Lagrangian
\eqref{Lagrangian}, did not appear in \eqref{KK_sum}. Let us
underline that only for $\varepsilon \gg 1$ we come to the
expression which can be obtained by neglecting widths of the massive
gravitons (\emph{zero width approximation}):%
\footnote{In fact, zero width approximation is justified if
inequality $s \gtrsim 3.5 \, \bar{M}_5$ is
satisfied~\cite{Kisselev:06}. Under this condition,
$\mathrm{Re}\,\mathcal{S}/\mathrm{Im}\,\mathcal{S} < 0.05$.}
\begin{equation}\label{KK_sum_zero_widths}
\mathrm{Im} \,\mathcal{S}(\hat{s}) \simeq - \frac{1}{2 \bar{M}_5^3
\sqrt{\hat{s}}} \;, \qquad  \mathrm{Re} \, \mathcal{S}(\hat{s})
\simeq 0 \;.
\end{equation}

In what follows, we will consider $\bar{M}_5 \geq 2(4)$ TeV for the
LHC energy $\sqrt{s} = 7(14)$ TeV. In such a case, the parameter
$\varepsilon$ in \eqref{parameters} obeys inequality $\varepsilon <
2$, and one must use the exact formulae \eqref{KK_sum},
\eqref{parameters} instead of the approximate expressions
\eqref{KK_sum_zero_widths}.%
\footnote{Since a mean value of the partonic energy $\sqrt{\hat{s}}$
is less than the collision energy $\sqrt{s}$, an effective value of
the parameter $\varepsilon$ is even less than 1.}

As one can see from \eqref{KK_sum}, the graviton spectrum in our
model consists of very sharp resonances whose widths are
proportional to $(\kappa/\bar{M}_5)^3$. The contribution from
\emph{one} resonance to the $p_{\perp}$-distribution can be
estimated as follows (more details can be found in
\cite{Kisselev:diphotons}):
\begin{equation}\label{one_res_cs}
\left. \frac{d \sigma (\mathrm{grav})}{d p_{\perp}} \right|_{one \
res} \sim \frac{\kappa}{\bar{M}_5^3 \sqrt{s}} \;.
\end{equation}
Since the total number of graviton resonances which contribute to
the cross section is proportional to $\sqrt{s}/\kappa$, we obtain at
fixed $x_{\perp}$
\begin{equation}\label{gravity_cs_1}
\frac{d \sigma (\mathrm{grav})}{d p_{\perp}} \sim
\frac{1}{\bar{M}_5^3} \;.
\end{equation}

Let us stress that in our scheme the gravity cross sections \emph{do
not depend} on the curvature $\kappa$, provided $\kappa \ll \bar{M}_5$,%
\footnote{Up to very small corrections of the type
$\mathrm{O}(\kappa/M_5)$.}
in contrast to the RS model with the large curvature in which
$\kappa \sim \bar{M}_5 \sim
\bar{M}_{\mathrm{Pl}}$~\cite{Randall:99}.

In order to obtain search limits for the LHC, we calculate
contributions from $s$-channel graviton resonances to the
$p_{\perp}$-distributions of the final muons for different values of
5-dimensional Planck scale $\bar{M}_5$. We use the MSTW 2008 NNLO
parton distributions~\cite{MSTW}, and convolute them with the
partonic cross sections (\ref{parton_cross_sec}) in
(\ref{cross_sec}). The PDF scale is taken to be equal to the
invariant mass of the muon pair, $\mu = M_{l^+l^-} =
\sqrt{\hat{s}}$.

We also impose the cut on the lepton pseudorapidities which is used
by ATLAS and CMS Collaborations in selecting dimuon
events~\cite{RS:dileptons_event_sel}:
\begin{equation}\label{rapidity_cut}
|\eta| < 2.4 \,.
\end{equation}
The limits on variations of the variables $\tau$ and $x_1$ in
\eqref{cross_sec}, resulting from this cut, are derived in
Appendix~A.

The reconstruction efficiency of 85$\%$ is assumed for the muon
events \cite{Efficiency}. Note that this assumption is not very
crucial for our scheme. For instance, given the efficiency varies
from 85$\%$ to 80$\%$, the LHC search limits on $\bar{M}_5$ diminish
by 1$\%$ only, due to the strong power-like dependence of the
gravity cross section on $\bar{M}_5$
\eqref{gravity_cs_1}.%
\footnote{Namely, a significant variation of $d\sigma
(\mathrm{grav})/dp_{\perp}$ can be achieved by a relatively small
variation of $\bar{M}_5$.}

\begin{figure}[hbtp]
 \begin{center}
    \resizebox{9cm}{!}{\includegraphics{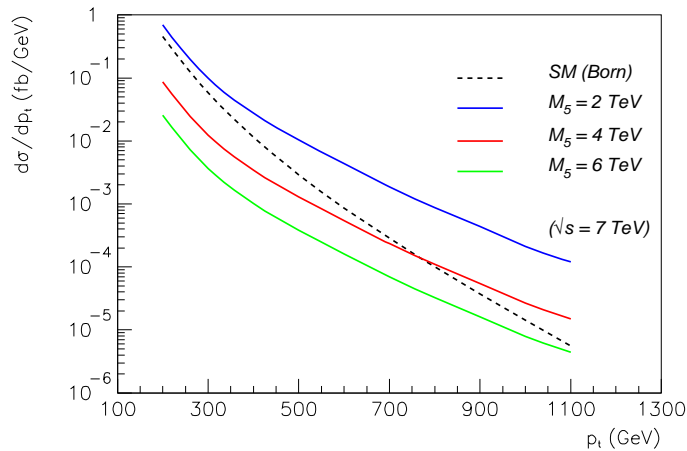}}
\caption{The KK graviton contribution to the dimuon production at
the LHC for several values of 5-dimensional \emph{reduced} Planck
scale (solid curves) vs. SM contribution (dashed curve) at $\sqrt{s}
= 7$ TeV.}
    \label{fig:DY_cs_7_TeV}
  \end{center}
\end{figure}

\begin{figure}[hbtp]
 \begin{center}
    \resizebox{9cm}{!}{\includegraphics{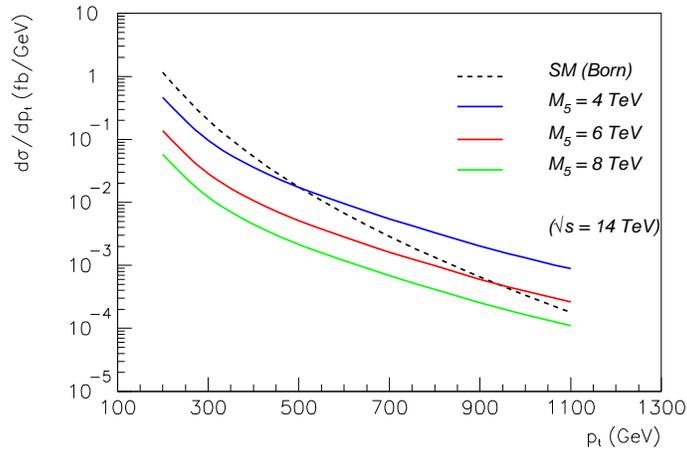}}
\caption{The same as in Fig.~\ref{fig:DY_cs_7_TeV}, but for the
energy $\sqrt{s} = 14$ TeV.}
    \label{fig:DY_cs_14_TeV}
  \end{center}
\end{figure}

\begin{figure}[hbtp]
 \begin{center}
    \resizebox{9cm}{!}{\includegraphics{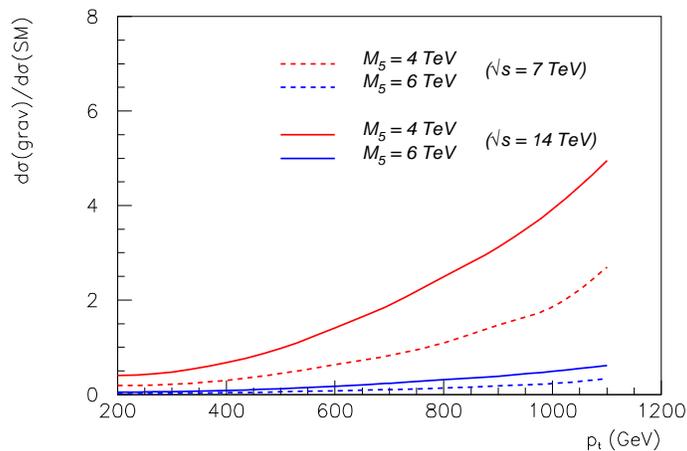}}
\caption{The ratio of the gravity induced cross sections to the SM
dimuon cross section at the LHC as a function of the muon transverse
momentum for several values of 5-dimensional \emph{reduced} Planck
scale.}
    \label{fig:DY_cs_ratio_TeV}
  \end{center}
\end{figure}

In Fig.~\ref{fig:DY_cs_7_TeV} and Fig.~\ref{fig:DY_cs_14_TeV} we
present the results of our calculations of gravity cross sections
for the dimuon production at the LHC. The gravity mediated
contributions to the cross sections do not include the SM
contribution.
The ratios of the gravity induced cross sections to the SM one is
shown in Fig.~\ref{fig:DY_cs_ratio_TeV}.

The differential cross section of the process under consideration is
represented in the form:
\begin{equation}\label{cross_sec_sum}
d\sigma = d \sigma (\mathrm{SM}) + d\sigma (\mathrm{grav}) + d\sigma
(\mathrm{SM}\!-\!\mathrm{grav}) \;,
\end{equation}
where the last term comes from the interference between the SM and
graviton interactions. Since the SM amplitude is pure real, while
the real part of each graviton resonance is antisymmetric with
respect to its central point, the interference term in
\eqref{cross_sec_sum} has appeared to be negligible in comparison
with the pure gravity contribution (the second term in
\eqref{cross_sec_sum}) \emph{after integration} in partonic momenta.
More details can be found in \cite{Kisselev:diphotons}.

The next Fig.~\ref{fig:DY_cs_14_TeV_zero} demonstrates us that an
ignorance of the graviton widths would be a \emph{rough
approximation}. As one can see, it results in very large suppression
of the cross sections. The reason of this suppression lies partially
in the fact that \cite{Kisselev:diphotons}
\begin{equation}\label{gravity_cs_2}
\frac{d \sigma (\mathrm{grav})}{d p_{\perp}} \sim
\frac{1}{p_{\perp}^3} \left( \frac{\sqrt{s}}{\bar{M}_5} \right)^3
\;,
\end{equation}
while in zero width approximation
\begin{equation}\label{gravity_cs_zero}
\left. \frac{d \sigma (\mathrm{grav})}{d p_{\perp}}
\right|_{\mathrm{zero \ width}} \sim \frac{1}{\bar{M}_5^3} \left(
\frac{\sqrt{s}}{\bar{M}_5} \right)^3 \;,
\end{equation}
as one can derive it from formulae \eqref{cross_sec},
\eqref{parton_cross_sec} and \eqref{KK_sum_zero_widths}.
\begin{figure}[hbtp]
 \begin{center}
    \resizebox{9cm}{!}{\includegraphics{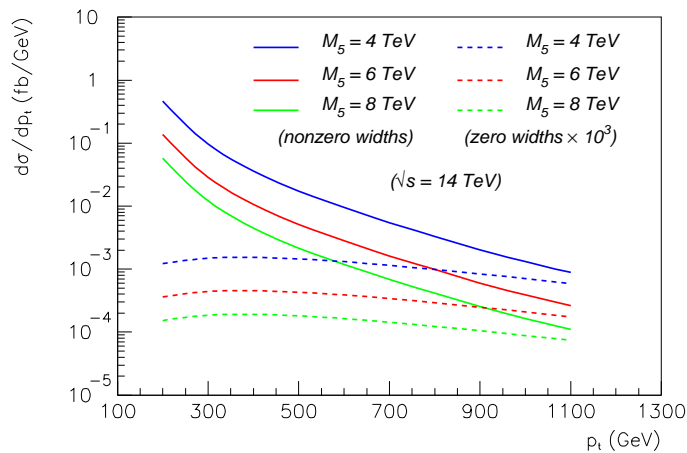}}
\caption{The contributions to the dimuon events from the KK
gravitons with nonzero widths (solid curves) vs. contributions from
zero width gravitons (multiplied by $10^3$, dashed curves) for the
different values of the \emph{reduced} 5-dimensional Planck scale.}
    \label{fig:DY_cs_14_TeV_zero}
  \end{center}
\end{figure}

\begin{figure}[hbtp]
 \begin{center}
    \resizebox{7cm}{!}{\includegraphics{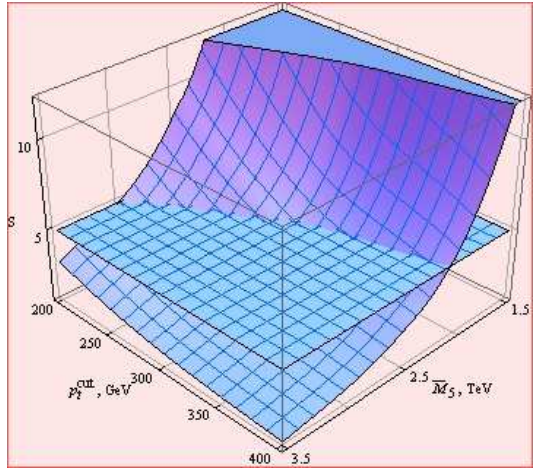}}
    \caption{The statistical significance $S$ for the dimuon
production at the LHC for $\sqrt{s} = 7$ TeV and integrated
luminosity 5 fb$^{-1}$ as a function of the transverse momentum cut
$p_{\perp}^{\mathrm{cut}}$ and \emph{reduced} 5-dimensional gravity
scale $\bar{M}_5$. The plane $S=5$ is also shown.}
    \label{fig:DY_S_7TeV_5fb}
  \end{center}
\end{figure}

\begin{figure}[hbtp]
 \begin{center}
    \resizebox{7cm}{!}{\includegraphics{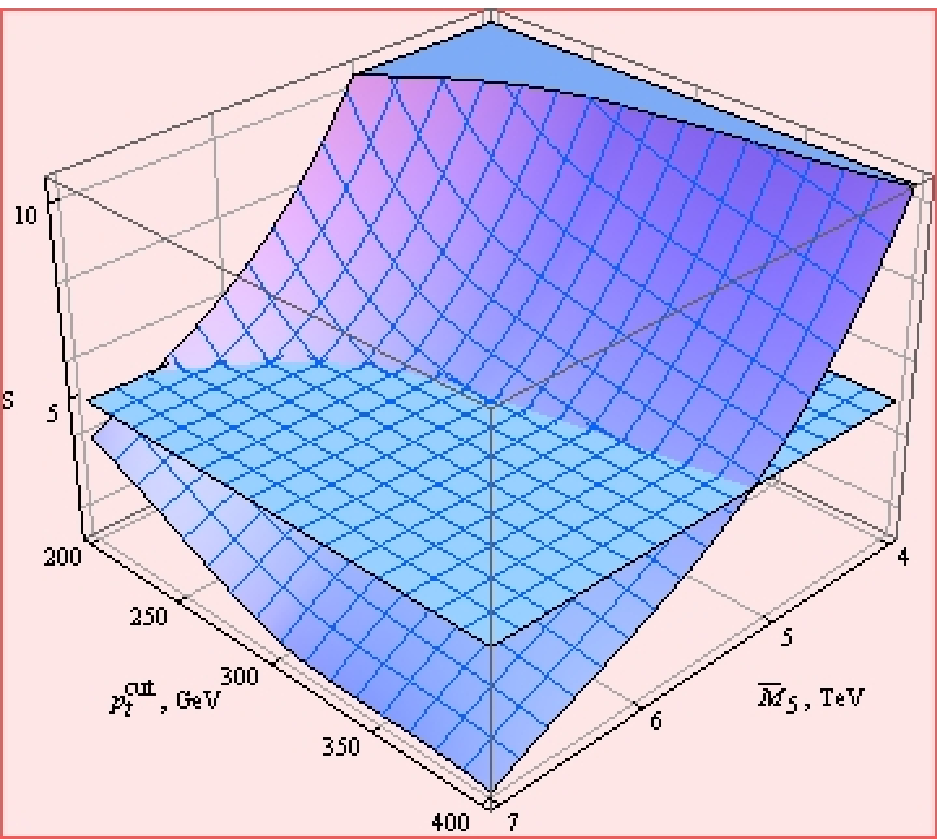}}
\caption{The same as in Fig.~\ref{fig:DY_S_7TeV_5fb}, but for
$\sqrt{s} = 14$ TeV and integrated luminosity 30 fb$^{-1}$.}
    \label{fig:DY_S_14TeV_30fb}
  \end{center}
\end{figure}

\begin{figure}[hbtp]
 \begin{center}
    \resizebox{7cm}{!}{\includegraphics{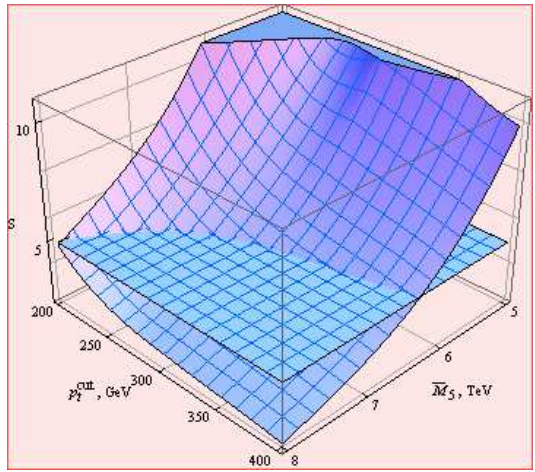}}
\caption{The same as in Fig.~\ref{fig:DY_S_7TeV_5fb}, but for
$\sqrt{s} = 14$ TeV and integrated luminosity 100 fb$^{-1}$.}
    \label{fig:DY_S_14TeV_100fb}
  \end{center}
\end{figure}

To take into account higher order contributions, we use a $K$-factor
1.5 for the SM background,%
\footnote{For the dimuon channel, the expected SM background
consists of $Z/\gamma^*$, $t\bar{t}$ and diboson events
($Z/\gamma^*$ being dominant), while ($W + \mathrm{jets}$) and QCD
background is negligible \cite{DY:background}.}
while a conservative value of $K=1$ is taken for the signal. Let
$N_S$($N_B$) be a number of signal (background) dimuon events with
$p_{\perp} > p_{\perp}^{\mathrm{cut}}$. Then we define the
statistical significance $\mathcal{S} = N_S/\sqrt{N_B}$, and require
a $5 \sigma$ effect. In Fig.~\ref{fig:DY_S_7TeV_5fb} the statistical
significance is shown for $\sqrt{s} = 7$ TeV as a function of the
transverse momentum cut $p_{\perp}^{\mathrm{cut}}$ and
\emph{reduced} 5-dimensional gravity scale $\bar{M}_5$.
Figs.~\ref{fig:DY_S_14TeV_30fb} and \ref{fig:DY_S_14TeV_100fb}
represent $\mathcal{S}$ for the dimuon events at $\sqrt{s} = 14$
TeV.

The aurthors of Ref.~\cite{Giudice:05} proposed a search in the
dilepton channel by cutting on the invariant masses $M_{l^+l^-}$
above 2 TeV. They found a 5 sigma bound on $M_5 = 4$ TeV (without
taking into account finite graviton widths). In our case, we have
the lower bound on $M_{l^+l^-} = 2p_{\perp}$ and integrate in
variable $\tau = M_{l^+l^-}^2/s$.



\section{Conclusions and discussions}
\label{sec3}

In the present paper the RS-like scenario with the small curvature
of the space-time \cite{Kisselev:05}-\cite{Kisselev:06} is
considered. In such a scheme, the reduced 5-dimensional Planck scale
$\bar{M}_5$ can vary from few TeV to tens TeV, while the curvature
$\kappa$ is allowed to vary from hundred MeV to few GeV (in fact,
the only condition $\kappa \ll \bar{M}_5$ should be satisfied). The
mass spectrum and experimental signature of the model are similar to
those in the ADD model~\cite{Arkani-Hamed:98} with one flat extra
dimension.

The $p_{\perp}$-distributions for the muon pairs production with
high $p_{\perp}$ at the LHC are calculated for the collision
energies $\sqrt{s} = 7$ TeV and $\sqrt{s} = 14$ TeV (see
Figs.~\ref{fig:DY_cs_7_TeV}-\ref{fig:DY_cs_ratio_TeV}). The
importance of the account of the KK graviton widths is demostrated
(see Fig.~\ref{fig:DY_cs_14_TeV_zero}). The statistical significance
as a function of the \emph{reduced} 5-dimensional Planck scale
$\bar{M}_5$ and cut on the lepton transverse momentum
$p_{\perp}^{\mathrm{cut}}$ is calculated (see
Figs.~\ref{fig:DY_S_7TeV_5fb}-\ref{fig:DY_S_14TeV_100fb}).

By using these results, we can obtain the following discovery limit
for the 7 TeV LHC and integrated luminosity of 5 fb$^{-1}$ in the
dimuon production:
\begin{equation}\label{search limit_7_TeV}
M_5 = 5.5 \mathrm{\ TeV} \;.
\end{equation}
Correspondingly, we obtain for the 14 TeV LHC:
\begin{equation}\label{search_limits_14_TeV}
M_5 = \left\{
\begin{array}{rl}
12.0 \mathrm{\ TeV} \;, & \mathcal{L} = 30 \mathrm{\
fb}^{-1} \\
14.6 \mathrm{\ TeV} \;,  & \mathcal{L} = 100 \mathrm{\
fb}^{-1} \\
\end{array}
\right.
\end{equation}
In deriving \eqref{search limit_7_TeV} and
\eqref{search_limits_14_TeV}, we used the relation $M_5 = (2
\pi)^{1/3}  \bar{M}_5$ (\ref{reduced_scale}).

It is important that these bounds on $M_5$ do not depend on the
curvature $\kappa$, contrary to the standard RS
model~\cite{Randall:99} in which estimated bounds on $M_5$ depend
significantly on the value of the ratio
$\kappa/\bar{M}_{\mathrm{Pl}}$.

Previously, analogous bounds were obtained for the diphoton
production at the Tevatron and 14 TeV LHC \cite{Kisselev:diphotons}.
Recently, results on the dilepton production at very high luminosity
(HL-LHC) were presented \cite{Kisselev:12}.

Our approach can be directly applied to the production of the
\emph{electron pairs} at the LHC. The only thing to do is to take
into account that for electron samples the transition region $1.37 <
|\eta| < 1.52$ ($1.44 < |\eta| < 1.57$) between the ECAL barrel and
endcap calorimeters is usually excluded in the ATLAS (CMS)
experiment, while the cuts $|\eta| < 2.47$ and $|\eta| < 2.5$ are
imposed by the ATLAS and CMS experiments, respectively (see, for
instance, \cite{RS:dileptons_event_sel}).



\section*{Acknowledgements}

The author is indebted to A.G.~Myagkov and S.V.~Shmatov for useful
discussions.



\begin{thebibliography}{11}
%
\bibitem{Randall:99}
L.~Randall and R.~Sundrum, Phys. Rev. Lett.  {\bf 83} (1999) 3370.
%
\bibitem{RS:Tevatron}
T.~Aaltonen \emph{et al.} (CDF Collaboration), Phys. Rev. Lett. {\bf
102} (2009) 091805; V.M.~Abazov \emph{et al.} (D0 Collaboration),
Phys. Rev. Lett. {\bf 104} (2010) 241802.
%
\bibitem{RS:dilepton_res}
G.~Aad \emph{et al.} (ATLAS Collaboration), Phys. Rev. Lett. {\bf
107} (2011) 272002; S.~Chatrchyan \emph{et al.} (CMS Collaboration),
Phys. Lett. B. {\bf 714} (2012) 158.
\bibitem{RS:diphoton_res}
G.~Aad \emph{et al.} (ATLAS Collaboration), Phys. Lett. B {\bf 70}
(2012) 538; S.~Chatrchyan \emph{et al.} (CMS Collaboration), Phys.
Rev. Lett. {\bf 108} (2012) 111801.
\bibitem{RS:ttbar_res}
G.~Aad \emph{et al.} (ATLAS Collaboration, Eur. Phys. J. C {\bf 72}
(2012) 2083.
\bibitem{RS:dileptons_event_sel}
S.~Viel (for the ATLAS Collaboration), \emph{Proc. of the XXXI
International Conference on Physics in Collision}, Vancouver, BC
Canada, August 28--September 1, 2011, arXiv:1111.2360; V.~Timciuc
(on behalf of the CMS Collaboration), \emph{Proc. of the XXXI
International Conference on Physics in Collision}, Vancouver, BC
Canada, August 28--September 1, 2011, arXiv:1111.4528.
\bibitem{RS:high_mass_res}
T.J.~Orimoto (on behalf of the CMS Collaboration), EPJ Web of
Conferences {\bf 28} (2012) 09010.
%
\bibitem{Kisselev:05}
A.V.~Kisselev and V.A.~Petrov, Phys. Rev. D {\bf 71} (2005) 124032.
\bibitem{Giudice:05}
G.~F.~Giudice, T.~Plehn and A.~Strumia, Nucl. Phys. B {\bf 706}
(2005) 455.
\bibitem{Kisselev:06}
A.V.~Kisselev, Phys. Rev. D {\bf 73} (2006) 024007.
%
\bibitem{Kisselev:ED}
A.V.~Kisselev, V.A.~Petrov and R.A.~Ryutin, Phys. Lett. B {\bf 630}
(2005) 100;
A.V.~Kisselev, JHEP {\bf 0703} (2007) 006.
%
\bibitem{Kisselev:diphotons}
A.V.~Kisselev, JHEP {\bf 0809} (2008) 039.
%
\bibitem{Boos:02}
E.E.~Boos, Yu.A.~Kubyshin, M.N.~Smolyakov and I.P.~Volobuev, Class.
Quan. Grav. {\bf 19} (2002) 4591; Nucl. Phys. B {\bf 717} (2005) 9.
%
\bibitem{Grinstein:01}
B.~Grinstain, D.R.~Nolte and W.~Skiba, Phys. Rev. D {\bf 63} (2001)
105005.
%
\bibitem{Arkani-Hamed:98}
N.~Arkani-Hamed, S.~Dimopoulos and G.~Dvali, Phys. Lett. B {\bf 429}
(1998) 263; I.~Antoniadis, N.~Arkani-Hamed, S.~Dimopoulos and
G.~Dvali, Phys. Lett. B {\bf 436} (1998) 257;  N.~Arkani-Hamed,
S.~Dimopoulos and G.~Dvali, Phys. Rev. D {\bf 59} (1999) 086004.
%
\bibitem{Matveev:69}
V.A.~Matveev, R.M.~Muradian and A.N.~Tavkhelidze, JINR P2-4543
(Dubna, 1969), SLAC TRANS-009: JINR R2-4543 (June, 1969); S.D.~Drell
and T.M.~Yan, SLAC-PUB-0755 (June, 1970), Phys. Rev. Lett. {\bf 25}
(1970) 316, \emph{errata} Phys. Rev. Lett. {\bf 25} (1970) 902.
%
\bibitem{DY:Tevatron}
T.~Affolder \emph{et al.} (CDF Collaboration), Phys. Rev. Lett. {\bf
84} (2000) 845. 10;  B.~Abbott \emph{et al.} (D0 Collaboration),
Phys. Rev. D {\bf 61} (2000) 032004,  Phys. Rev. Lett. {\bf 84}
(2000) 2792; V.M.~Abazov \emph{et al.} (D0 Collaboration), Phys.
Rev. Lett. {\bf 100} (2008) 102002, \emph{ibid} {\bf 104} (2010)
241802.
%
\bibitem{LHC_limits:ADD}
G.~Aad \emph{et al.} (ATLAS Collaboration), Phys. Lett. B {\bf 716}
(2012) 122; S.~Chatrchyan \emph{et al.} (CMS Collaboration), JHEP
{\bf 05} (2011) 085, Phys. Lett. B {\bf 711} (2012) 15.
%
\bibitem{Davoudiasl:00}
H.~Davoudiasl, J.L.~Hewett and T.G.~Rizzo, Phys. Rev. Lett. {\bf 84}
(2000) 2080; Phys. Rev. D {\bf 63} (2001) 075004.
%
\bibitem{Osland:08}
P.~Osland, A.A.~Pankov, A.V.~Tsytrinov and N.~Paver, Phys. Rev. D
{\bf 78} (2008) 035008.
%
\bibitem{MSTW}
A.D.~Martin, W.J.~Stirling, R.S.~Thorne and G.~Watt, Eur. Phys. J. C
\textbf{63} (2009) 189.
%
\bibitem{Efficiency}
G.~Aad \emph{et al.} (ATLAS Collaboration), arXiv:1209.2535.
%
\bibitem{DY:background}
The ATLAS Collaboration, Proc. of the \emph{47-th Rencontres de
Moriond on Electroweak Interactions and Unified Theories}, La
Thuile, Italy, 3-10 March 2012, arXiv:1111.2360.
%
\bibitem{Kisselev:12}
A.V.~Kisselev, \emph{Talk presented at the CMS Workshop on
Perspectives on Physics and on CMS at Very High Luminosity},
Alushta, Crimea, Ukraine, 28--31 May, 2012, arXiv:1208.3844.
%
\end{thebibliography}



\setcounter{equation}{0}
\renewcommand{\theequation}{A.\arabic{equation}}

\section*{Appendix A}
\label{app:A}

In this section we will calculate the integration region in
\eqref{cross_sec} in a case when a cut on final lepton
(pseudo)rapidities is imposed. Let us consider a production of a
lepton pair with high transverse momenta $p_{\perp}$ in $pp$
collision:
\begin{equation}\label{dilepton_pp}
p p \rightarrow l^+ \, l^- + X \;, \quad l = e, \mu \;.
\end{equation}
In the c.m.s. of the initial protons, their 4-momenta are
\begin{equation}\label{proton_mom_lab}
p_{1,2}^{\mu} = \frac{\sqrt{s}}{2} (1, \ \vec{0}, \ \pm 1) \;,
\end{equation}
where $\sqrt{s}$ is the collision energy. Let us call this system
\emph{lab-system}, contrary to the c.m.s. of the final leptons. In
what follows, we neglect the masses of the protons and leptons with
respect to $\sqrt{s}$, as well as inner transverse momenta of the
partons inside the protons with respect to $p_{\perp}$.

Let $x_1$, $x_2$ be momentum fractions of the partons $q_1$, $q_2$
inside the colliding protons. Then the 4-momenta of the final
leptons are:
\begin{equation}\label{lepton_mom}
l_{1,2}^{\mu} = (l_{1,2}^0, \ \pm \, \vec{p}_{\perp}, \
l_{1,2}^{\parallel}) \;,
\end{equation}
with
\begin{align}\label{lepton_mom_lab}
l_{1,2}^{\parallel} &= \frac{\sqrt{s}}{4} \left[ (x_1 - x_2) \pm
(x_1 + x_2)\sqrt{1 - \frac{x_{\perp}^2}{\tau}} \right] \;,
\nonumber \\
l_{1,2}^0 &= \frac{\sqrt{s}}{4} \left[ (x_1 + x_2) \pm (x_1 - x_2)
\sqrt{1 - \frac{x_{\perp}^2}{\tau}} \right] \;.
\end{align}
The dimensional variables $\tau$ and $x_{\perp}$ were defined in the
main text \eqref{tau_xtr}. They obey kinematical inequalities
\eqref{int_region_full}. The invariant mass of the lepton pair is
$\hat{s} = s \, x_1 x_2$.

In the c.m.s. of the final leptons (herein called \emph{cm-system}),
the proton momenta look like
\begin{align}\label{proton_mom_cm}
p_1^{\mu} &= \frac{\sqrt{s}}{2} \, \left( \sqrt{\frac{x_2}{x_1}}, \
\vec{0}, \  \sqrt{\frac{x_2}{x_1}} \right) \, ,
\nonumber  \\
p_2^{\mu} &= \frac{\sqrt{s}}{2} \, \left( \sqrt{\frac{x_1}{x_2}}, \
\vec{0}, \ - \sqrt{\frac{x_1}{x_2}} \right) \;,
\end{align}
while the parton momenta are
\begin{equation}\label{parton_mom_cm}
q_{1,2}^{\mu} = \frac{\sqrt{s}}{2} \, (\sqrt{x_1 x_2}, \ \vec{0} , \
\pm \, \sqrt{x_1 x_2}) \;.
\end{equation}
Correspondingly, the momenta of the leptons looks like
\begin{equation}\label{lepton_mom_lab}
l_{1,2}^{\mu}= \left[ \frac{\sqrt{s \tau}}{2}, \ \pm \,
\vec{p}_{\perp}, \ \pm \frac{\sqrt{s \tau}}{2} \, \sqrt{1 -
\frac{x_{\perp}^2}{\tau}} \; \right] \;.
\end{equation}
Note, since the lab-system and cm-system are related by a Lorentz
boost along the beam axis, the partons momentum fractions remains
the same in both systems.

Let us turn to the lab-system. The rapidity of the lepton pair in
this system is equal to $(1/2)\ln (x_1/x_2)$. This means that the
lepton rapidities are given by
\begin{align}\label{rapidities_lab}
y_1 &= \ln \left[ \frac{x_1}{x_{\perp}} \left( 1  + \sqrt{1 -
\frac{x_{\perp}^2}{\tau}} \right) \right]  \;,
\nonumber \\
y_2 &= - \ln \left[ \frac{x_2}{x_{\perp}} \left( 1 + \sqrt{1 -
\frac{x_{\perp}^2}{\tau}} \right) \right] \;.
\end{align}

We are interested in the production of the leptons with the high
momenta ($p_{\perp} \geq 200$ GeV). That is why, in order to
estimate the integration region in \eqref{cross_sec}, one can safely
impose a cut on lepton \emph{rapidities} $y_{1,2}$, instead of
using cut on their \emph{pseudorapidities}:%
\footnote{Remember that the rapidity is defined as $y =
(1/2)\ln[(E+p_z)/(E-p_z)]$, while the psedorapidity $\eta =
(1/2)\ln[(p+p_z)/(p-p_z)] = -\ln[\tan(\theta/2)]$. For a
relativistic particle ($p \gg m$), one gets $y \simeq \eta$ for
$\theta \gg m/p$.}
\begin{equation}\label{y_cut}
\mid y_{1,2} \mid \leq y_{\mathrm{cut}} \;.
\end{equation}
This bound is equivalent to the following inequalities:
\begin{equation}\label{x1_cut}
A^{-1} \leq \frac{x_1}{\sqrt{\tau}} \leq A \;,
\end{equation}
where
\begin{equation}\label{A}
A = \frac{x_{\perp} \exp(y_{\mathrm{cut}})}{\sqrt{\tau} + \sqrt{\tau
- x_{\perp}^2}} \;.
\end{equation}
In order to get a non-zero integration region for $x_1$, one must
take $A \geq 1$. It leads to
\begin{equation}\label{tau_cut}
\tau \leq x_{\perp}^2 \cosh^2 \! y_{\mathrm{cut}} \;.
\end{equation}

The inequalities \eqref{x1_cut} and \eqref{tau_cut} must be treated
simultaneously with the inequalities \eqref{int_region_full}.
Depending on values of $x_{\perp}$, the integration region splits
into several parts:
\begin{enumerate}
\item
$0 \leq x_{\perp} \leq \exp(-y_{\mathrm{cut}})$. In this case:
\begin{align} \label{int_region_pt_1}
x_{\perp}^2 &\leq \tau \leq  x_{\perp}^2 \cosh^2 \! y_{\mathrm{cut}}
\, ,
\nonumber \\
\sqrt{\tau} A^{-1} & \leq x_1 \leq \ \sqrt{\tau} A \;.
\end{align}
\item
$\exp(-y_{\mathrm{cut}}) < x_{\perp} \leq (\cosh
y_{\mathrm{cut}})^{-1}$. Variables $\tau$, $x_1$ run two subregions:
\begin{align}\label{int_region_pt_2_1}
x_{\perp}^2 &\leq \tau < \tau_0  \, ,
\nonumber \\
\tau &\leq x_1 \leq 1 \;,
\end{align}
and
\begin{align}\label{int_region_pt_2_2}
\tau_0 &\leq \tau \leq x_{\perp}^2 \cosh^2 \! y_{\mathrm{cut}} \, ,
\nonumber \\
\sqrt{\tau} A^{-1} & \leq x_1 \leq \ \sqrt{\tau} A \;.
\end{align}
Here we introduced the notation:%
\footnote{Note that $x_{\perp}^2 \leq \tau_0 \leq x_{\perp}^2
\cosh^2 \! y_{\mathrm{cut}}$ for the values of $x_{\perp}$ under
consideration.}
\begin{equation}\label{tau0}
\tau_0 = \frac{x_{\perp} \exp (-y_{\mathrm{cut}})}{2 - x_{\perp}
\exp (y_{\mathrm{cut}})} \;.
\end{equation}
\item
$(\cosh y_{\mathrm{cut}})^{-1} < x_{\perp} \leq 1$. In this case,
variables $\tau$, $x_1$ run the region \eqref{int_region_full}.
\end{enumerate}

Without cuts on rapiditites (formally, in the limit
$y_{\mathrm{cut}} \rightarrow \infty$), the full kinematically
allowed region \eqref{int_region_full} is restored.

\end{document}